# A note on some perfect fluid Kantowski-Sachs and Bianchi type III space-times and their conformal vector fields in f(R) theory of gravity

Ghulam Shabbir[1,2], Fiaz Hussain[3], A. H. Kara[4] and Muhammad Ramzan[3]

[1] Faculty of Engineering Sciences, GIK Institute of Engineering Sciences and Technology, Topi, Swabi, KPK, Pakistan. [2]shabbir@giki.edu.pk

[3] Department of Mathematics, The Islamia University of Bahawalpur, Pakistan.

[4] School of Mathematics and DST-NRF Center of Excellence in Mathematical and Statistical Sciences, University of the Witwatersrand, Johannesburg, Wits 2050, South Africa.

## Abstract

The purpose of this paper is to find conformal vector fields of some perfect fluid Kantowski-Sachs and Bianchi type III space-times in the f(R) theory of gravity using direct integration technique. In this study there exists only eight cases. Studying each case in detail, we found that in two cases proper conformal vector fields exist while in the rest of six cases conformal vector fields become Killing vector fields. The dimension of conformal vector fields is either 4 or 6.

## 1 Introduction

General theory of relativity is based upon the Einstein field equations (EFEs), which relates the geometry and physics of space-time. These equations are highly non-linear, therefore it is very difficult to find their exact solutions. One of the ways to find the solutions of EFEs is to impose some symmetry restrictions. These symmetry restrictions have their own place in Einstein theory of general relativity as these are found to be extremely helpful in finding the exact solutions of EFEs. Some of the most basic symmetries are Killing symmetry, in which the Lie derivative of metric tensor vanishes, homothetic symmetry, in which the Lie derivative of metric tensor is invariant upto a constant multiple, conformal symmetry, in which the Lie derivative of metric tensor is invariant upto a function. Using these symmetries one can obtain certain conservation laws. These conservation laws play a vital role in physics as the invariance of physical quantities under a given transformation help physicists to simplify their problems. For example the existence of time like symmetry (also known as time-like Killing vector field) ensures conservation of energy of a particle defined by the equation $E = vp$, where $v$ and $p$ represents time-like Killing vector field and momentum of the particle respectively [1]. Moreover, symmetry restrictions are found to be extremely useful for the investigation of gravitational waves as well as to address the problem of localization of energy momentum in general relativity (GR). Likewise, in the case of spherical symmetric space-time the rotational symmetries provide conservation of angular momentum [2]. Geometrically, a space-time depicting repetitive structure at different scales is self-similar due to the existence of homothetic symmetry. On the other hand, in conformal symmetry distance measured between two points is scaled by a conformal function with the property that the angle between two curves always remain the same [3]. Thus, the conformal symmetries are transformations which preserve the structure of geometrical objects. With the help of conformal symmetry one can obtain conformal vector fields which has wide range of applications at geometric, kinematics and dynamic level. Some more applications of conformal vector fields are given in [4-6]. A reasonable amount of work has been done on classification of different space-times according to their conformal vector fields [7-9].

There is no doubt that the general theory of relativity is an elegant theory of gravitation to describe the structure of space-time gravity and matter. Recently astronomical data shows that our universe has undergone a phase of accelerated expansion [10, 11]. It is expected that this expansion might be due to some unknown form of energy called dark energy. An alternative for dark energy is assumed to be a cosmological constant [12]. To tackle such type of issues a number of modifications in general relativity (GR) have been made. The f(R) theory of gravity is one of the modifications of GR in which Ricci scalar $R$ is replaced by a function f(R) in the standard Einstein Hilbert action [13]. Due to successful aspects of this theory, a lot of attention has been given in the last couple of years [14-19]. On the other hand, in the f(R) theory of gravity, f(R) models play important role in cosmology as they help to justify late-time acceleration and early time inflation [20]. For a detail review of these f(R) models, we left the reader to [21-23]. These f(R) models also help to explore solutions of EFEs with constant and non-constant curvature. For the investigation of the physical properties of specific models the exact solutions of EFEs play a pivotal role. Moreover, by classifying a space-time according to some symmetry restriction makes easier to find solutions of EFEs. In 1966, Kantowski and Sachs [24] studied some spatially homogeneous anisotropic relativistic cosmological models and discussed dust matter solutions of EFEs. After that a series of work on Kantowski-Sachs and Bianchi type III space-times get started. A large body of literature is devoted to discuss Kantowski-Sachs and Bianchi type III space-times according to different symmetries in GR and teleparallel theory of gravitation [25-32].

Over the past few decades exploring solutions in extended theories of gravity especially in the f(R) theory of gravity has remained a key subject for the researchers. A brief picture of work in this era is given here, Taşer, [33] found some Bianchi type I cosmological models with the help of conformal symmetry in the f(R) theory of gravity. Same authors [34] discussed Chaplygin gas solutions of static spherically symmetric space-times via conformal symmetry in the f (R) theory of gravity. Samanta [35] studied the Kantowski-Sachs universe filled with perfect fluid in the f(R,T) theory of gravity. Shamir [36] discussed some vacuum Kantowski-Sachs and Bianchi type III cosmological models in the f(R) theory of gravity. Very recently, Hussain et al, [37] explored some perfect fluid solutions of static cylindrically symmetric space-times in the f(R) theory of gravity and found conformal vector fields of resulting solutions. Shabbir et al, [38] classified spherically symmetric static space-times in the f(R) theory of gravity according to their conformal vector fields. Shabbir et al, [39] also classified Bianchi type I space-times according to their conformal vector fields in the f(R) theory of gravity. Our purpose in this paper is to find conformal vector fields of some perfect fluid Kantowski-Sachs and Bianchi type III space-times in the f(R) theory of gravity. A conformal vector field $X$ is defined by [40]

$$L_X g_{ab} \equiv g_{ab,c} X^c + g_{bc} X^c_{,a} + g_{ac} X^c_{,b} = 2\psi g_{ab}, \tag{1}$$

where $\psi$ is the smooth conformal function. Here, $L$, $g_{ab}$ and comma denotes the Lie derivative, metric tensor and partial derivative, respectively. If $\psi$ is a constant, then $X$ represents a homothetic vector field (proper homothetic if $\psi \neq 0$) and if it is zero, then $X$ becomes a Killing vector field. If the vector field $X$ is conformal but not homothetic then it is called proper conformal vector field. Here, it is important to mention for the readers that a different approach for obtaining a related symmetry classification can be found in [41-42].

## 2 Main Results

Consider Kantowski-Sachs and Bianchi type III space-times in the usual coordinates $(t, r, \theta, \phi)$ (given by $(x^0, x^1, x^2, x^3)$ respectively) with the line element [2]

$$ds^2 = -dt^2 + A dr^2 + B\left[d\theta^2 + l(\theta)^2 d\phi^2\right], \tag{2}$$

where $A = A(t)$ and $B = B(t)$ are nowhere zero functions of $t$ only. For $l(\theta) = Sin\theta$ the above space-times (2) become Kantowski-Sachs space-times and for $l(\theta) = Sinh\theta,$ the above space-times (2) become Bianchi type III space-times. The above space-time (2) admit four linearly independent Killing vector fields which are [32]

$$\frac{\partial}{\partial r},\ \frac{\partial}{\partial \phi},\ \cos\phi\frac{\partial}{\partial \theta}-\frac{l'}{l}\sin\phi\frac{\partial}{\partial \phi},\ \sin\phi\frac{\partial}{\partial \theta}+\frac{l'}{l}\cos\phi\frac{\partial}{\partial \phi}, \tag{3}$$

where prime denotes the derivative with respect to $\theta.$ Ricci scalar $R$ for the space-times (2) is

$$R = \frac{1}{2}\left[\frac{\dot{A}^2}{A^2}-\frac{2\dot{A}\dot{B}}{AB}+\frac{\dot{B}^2}{B^2}-\frac{2\ddot{A}}{A}-\frac{4\ddot{B}}{B}+\frac{4\alpha}{B}\right], \tag{4}$$

where $\alpha = \frac{l''}{l}$ and dot denotes the derivative with respect to $t$. Using equation (2) in equation (1), we have

$$X^0_{,0} = \psi, \tag{5}$$

$$X^0_{,1} - AX^1_{,0} = 0, \tag{6}$$

$$X^0_{,2} - BX^2_{,0} = 0, \tag{7}$$

$$X^0_{,3} - Bl^2X^3_{,0} = 0, \tag{8}$$

$$\dot{A}X^0 + 2AX^1_{,1} = 2A\psi, \tag{9}$$

$$AX^1_{,2} + BX^2_{,1} = 0, \tag{10}$$

$$AX^1_{,3} + Bl^2X^3_{,1} = 0, \tag{11}$$

$$\dot{B}X^0 + 2BX^2_{,2} = 2B\psi, \tag{12}$$

$$X^2_{,3} + l^2X^3_{,2} = 0, \tag{13}$$

$$\dot{B}\,l\,X^0 + 2Bl'X^2 + 2Bl\,X^3_{,3} = 2Bl\psi. \tag{14}$$

Equation (5) gives $X^0 = \int \psi dt + S^1(r,\theta,\phi),$ where $S^1(r,\theta,\phi)$ is a function of integration. Substituting the value of $X^0$ in equations (6), (7) and (8), we have

$$\left.\begin{aligned}
&X^0 = \int \psi dt + S^1(r,\theta,\phi),\ X^1 = \int\left(\frac{1}{A}\int \psi_{,r} dt\right)dt + S^1_{r}(r,\theta,\phi)\int\frac{dt}{A} + S^2(r,\theta,\phi),\\
&X^2 = \int\left(\frac{1}{B}\int \psi_{\theta} dt\right)dt + S^1_{\theta}(r,\theta,\phi)\int\frac{dt}{B} + S^3(r,\theta,\phi),\\
&X^3 = \frac{1}{l^2}\left[\int\left(\frac{1}{B}\int \psi_{\phi}\, dt\right)dt + S^1_{\phi}(r,\theta,\phi)\int\frac{dt}{B}\right] + S^4(r,\theta,\phi),
\end{aligned}\right\} \qquad (15)$$

where $S^2(r,\theta,\phi)$, $S^3(r,\theta,\phi)$ and $S^4(r,\theta,\phi)$ are functions of integration. It is important to mention here that in this paper, we are interested in finding the conformal vector field $X$ for the perfect fluid Kantowski-Sachs and Bianchi type III space-times in the f(R) theory of gravity. Therefore, first we need to explore solutions in the f(R) theory of gravity. For this, we start with the EFEs in the f(R) theory of gravity [36]

$$F(R)R_{ab} - \frac{1}{2}f(R)g_{ab} - \nabla_a\nabla_b F(R) + g_{ab}\Box F(R) = kT_{ab}, \qquad (16)$$

where $f(R)$ is the function of Ricci scalar $R$, $F(R) \equiv \frac{d}{dR}f(R)$, $k$ is the coupling constant, $T_{ab}$ is the energy momentum tensor and $\Box \equiv \nabla^a\nabla_a$ in which $\nabla_a$ is the covariant derivative. The source of energy momentum tensor is perfect fluid which is

$$T_{ab} = (\rho + p)u_a u_b + p g_{ab}, \qquad (17)$$

where $\rho$ and $p$ are the matter density and pressure of the fluid respectively. Here, $u_a$ is the four velocity vector defined as $u_a = -\delta^0_a$. Using equations (2) and (17) in equation (16) and after some straightforward calculations, we get

$$\ddot{F} + \left(\frac{\dot{A}}{2A} + \frac{2\dot{B}}{B}\right)\dot{F} + \left(\frac{\dot{A}\dot{B}}{2AB} + \frac{\ddot{B}}{B} - \frac{2\alpha}{B} - R\right)F + f = k(\rho - p). \qquad (18)$$

$$\ddot{F} + \left(\frac{\dot{A}}{A} + \frac{3\dot{B}}{2B}\right)\dot{F} + \left(\frac{3\dot{A}\dot{B}}{4AB} + \frac{\ddot{B}}{2B} + \frac{\ddot{A}}{2A} - \frac{\dot{A}^2}{4A^2} - \frac{\alpha}{B} - R\right)F + f = k(\rho - p). \qquad (19)$$

Subtracting equation (19) from equation (18), gives

$$\left(\frac{\dot{A}}{A} - \frac{\dot{B}}{B}\right)\dot{F} + \left(\frac{\ddot{A}}{A} - \frac{\ddot{B}}{B} + \frac{\dot{A}\dot{B}}{2AB} - \frac{\dot{A}^2}{2A^2} + \frac{2\alpha}{B}\right)F = 0. \qquad (20)$$

Now, our purpose is to obtain solution of equation (20). Here, we will skip all the details and only present the results which are:

(i) $A=(c_1t+c_2)$, $B=(c_1t+c_2)^{1/2}$, $R=\frac{1}{2}\left[\frac{4\alpha}{\sqrt{(c_1t+c_2)}}+\frac{5c_1^2}{4(c_1t+c_2)^2}\right]$ and

$f(R)=\mathrm{e}^{\left(-\frac{8\alpha}{3c_1^2}(c_1t+c_2)^{3/2}+c_3\right)}R+c_4$, where $c_1,c_2,c_3,c_4\in\Re\,(c_1\neq 0)$.

(ii) $A=4(c_1t+c_2)^{-2}$, $B=\frac{1}{4}(c_1t+c_2)^2$, $R=2\left[\frac{4\alpha-c_1^2}{(c_1t+c_2)^2}\right]$ and $f(R)=\left[c_3(c_1t+c_2)^{2\alpha/c_1^2}\right]R+c_4$,

where $c_1,c_2,c_3,c_4\in\Re\,(c_1\neq 0\,\text{and}\,c_3\neq 0)$.

(iii) $A=(2c_1t+2c_2)^{1/2}$, $B=(2c_1t+2c_2)^{3/4}$, $R=\frac{1}{32}\left[\frac{16(2c_1t+2c_2)^{5/4}\alpha+21c_1^2}{(c_1t+c_2)^2}\right]$ and

$f(R)=\mathrm{e}^{\left(\frac{8\alpha}{5c_1^2}(2c_1t+2c_2)^{5/4}+c_3\right)}R+c_4$, where $c_1,c_2,c_3,c_4\in\Re\,(c_1\neq 0)$.

(iv) $A=\left(\frac{c_1t+c_2}{3}\right)^3$, $B=\left(\frac{c_1t+c_2}{3}\right)^{-1/2}$, $R=\frac{1}{24}\left[16\sqrt{3c_1t+3c_2}-\frac{33c_1^2}{(c_1t+c_2)^2}\right]$ and

$f(R)=\mathrm{e}^{\left(\frac{-8\alpha}{35\sqrt{3}c_1^2}(c_1t+c_2)^{5/2}+c_3\right)}R+c_4$, where $c_1,c_2,c_3,c_4\in\Re\,(c_1\neq 0)$.

(v) $A=\text{constant}$, $B=(\alpha t^2+c_1t+c_2)$, $R=\frac{1}{2}\left[\frac{c_1^2-4\alpha c_2}{(\alpha t^2+c_1t+c_2)^2}\right]$ and $f(R)=c_3R+c_4$, where

$c_1,c_2,c_3,c_4\in\Re\,(c_1\neq 0\,\text{and}\,c_3\neq 0)$.

(vi) $A=(c_1t+c_2)^2$, $B=\text{constant}$, $R=2\alpha$ and $f(R)=\mathrm{e}^{\left[\frac{-\alpha}{c_1}\left(c_1\frac{t^2}{2}+c_2t\right)+c_3\right]}R+c_4$, where

$c_1,c_2,c_3,c_4\in\Re\,(c_1\neq 0)$.

(vii) $A=t^4$, $B=t^{-1}$, $R=\frac{1}{2}\left[\frac{4\alpha t^3-7}{t^2}\right]$ and $f(R)=c_1\,\mathrm{e}^{\frac{-2\alpha t^3}{15}}R+c_2$, where $c_1,c_2\in\Re\,(c_1\neq 0)$.

(viii) $A=\text{constant}$, $B=t^2$, $R=2\left[\frac{\alpha-1}{t^2}\right]$ and $f(R)=c_1t^{\alpha-1}R+c_2$, where $c_1,c_2\in\Re\,(c_1\neq 0)$.

We will discuss each case in turn.

## Case (i):

In this case we have $A=(c_1t+c_2)$, $B=(c_1t+c_2)^{1/2}$, $R=\frac{1}{2}\left[\frac{4\alpha}{\sqrt{(c_1t+c_2)}}+\frac{5c_1^2}{4(c_1t+c_2)^2}\right]$ and

$f(R)=e^{\left(-\frac{8\alpha}{3c_1^2}(c_1t+c_2)^{3/2}+c_3\right)}R+c_4,$ where $c_1,c_2,c_3,c_4\in\Re\,(c_1\neq 0)$. The space-time (2) takes the form:

$$ds^2=-dt^2+(c_1t+c_2)dr^2+(c_1t+c_2)^{1/2}\left[d\theta^2+l(\theta)^2d\phi^2\right]. \tag{21}$$

Now we find conformal vector fields of the space-time (21) using equations (5) to (14). If one proceeds further, after some lengthy calculations, one finds that $\psi=0$, which means that no proper conformal vector fields exist. Here conformal vector fields are the Killing vector fields which are given in equation (3).

## Case (ii):

Here we have $A=4(c_1t+c_2)^{-2}$, $B=\frac{1}{4}(c_1t+c_2)^2$, $R=2\left[\frac{4\alpha-c_1^2}{(c_1t+c_2)^2}\right]$ and

$f(R)=\left[c_3(c_1t+c_2)^{2\alpha/c_1^2}\right]R+c_4,$ where $c_1,c_2,c_3,c_4\in\Re\,(c_1\neq 0\,\text{and}\,c_3\neq 0)$. The space-time (2) becomes

$$ds^2=-dt^2+4(c_1t+c_2)^{-2}dr^2+\frac{1}{4}(c_1t+c_2)^2\left[d\theta^2+l(\theta)^2d\phi^2\right]. \tag{22}$$

Again solving equations (5) to (14) with the help of equation (22), one has following components of conformal vector fields:

$$\left.\begin{aligned}&X^0=(c_1t+c_2)(c_5r+c_6),\ X^1=2c_1\left[\left[\left(\frac{16c_1^2r^2+(c_1t+c_2)^4}{32c_1^2}\right)c_5+c_6r\right]+c_7,\right.\\&X^2=c_8\cos\phi+c_9\sin\phi,\ X^3=\frac{l'}{l}(-c_8\sin\phi+c_9\cos\phi)+c_{10}\end{aligned}\right\}, \tag{23}$$

with conformal factor $\psi=c_1(c_5r+c_6)$, where $c_5,c_6,c_7,c_8,c_9,c_{10}\in\Re$. The above space-time (22) admits six conformal vector fields in which four are Killing vector fields which are given in equation (3), one is proper homothetic vector field which is $t\frac{\partial}{\partial t}+2r\frac{\partial}{\partial r}$ and one is proper conformal vector field. Proper conformal vector fields after subtracting the homothetic vector fields from equation (23) are

$$X^0 = c_5 r(c_1 t + c_2) + c_{11}, X^1 = \left(\frac{16c_1^2 r^2 + (c_1 t + c_2)^4}{16c_1}\right) c_5, X^2 = 0, X^3 = 0, \qquad (24)$$

where $c_{11} = c_2 c_6$.

## Case (iii):

In this case one has $A = (2c_1 t + 2c_2)^{1/2}$, $B = (2c_1 t + 2c_2)^{3/4}$, $R = \frac{1}{32}\left[\frac{16(2c_1 t + 2c_2)^{5/4}\alpha + 21c_1^2}{(c_1 t + c_2)^2}\right]$ and

$f(R) = \mathrm{e}^{\left(\frac{8\alpha}{5c_1^2}(2c_1 t + 2c_2)^{5/4} + c_3\right)} R + c_4$, where $c_1, c_2, c_3, c_4 \in \Re\,(c_1 \neq 0)$. The space-time (2) takes the form:

$$ds^2 = -dt^2 + (2c_1 t + 2c_2)^{1/2} dr^2 + (2c_1 t + 2c_2)^{3/4}\left[d\theta^2 + l(\theta)^2 d\phi^2\right]. \qquad (25)$$

Adopting the same procedure as in the previous case, we get $\psi = 0 \Rightarrow$ conformal vector fields in this case are Killing vector fields which are given in equation (3).

## Case (iv):

Now, if $A = \left(\frac{c_1 t + c_2}{3}\right)^3$, $B = \left(\frac{c_1 t + c_2}{3}\right)^{-1/2}$, $R = \frac{1}{24}\left[16\sqrt{3c_1 t + 3c_2} - \frac{33c_1^2}{(c_1 t + c_2)^2}\right]$ and

$f(R) = \mathrm{e}^{\left(\frac{-8\alpha}{35\sqrt{3}c_1^2}(c_1 t + c_2)^{5/2} + c_3\right)} R + c_4$, where $c_1, c_2, c_3, c_4 \in \Re\,(c_1 \neq 0)$. the space-time (2) becomes:

$$ds^2 = -dt^2 + \left(\frac{c_1 t + c_2}{3}\right)^3 dr^2 + \left(\frac{c_1 t + c_2}{3}\right)^{-1/2}\left[d\theta^2 + l(\theta)^2 d\phi^2\right]. \qquad (26)$$

Solving equations (5) to (14) with the help of space-time (26) again we get $\psi = 0$ which implies that conformal vector fields in this case are Killing vector fields which are given in equation (3).

## Case (v):

Here $A = \text{constant}$, $B = (\alpha t^2 + c_1 t + c_2)$, $R = \frac{1}{2}\left[\frac{c_1^2 - 4\alpha c_2}{(\alpha t^2 + c_1 t + c_2)^2}\right]$ and $f(R) = c_3 R + c_4$, where $c_1, c_2, c_3, c_4 \in \Re\,(c_1 \neq 0 \text{ and } c_3 \neq 0)$. The space-time (2) after suitable rescaling of $r$ takes the form

$$ds^2 = -dt^2 + dr^2 + (\alpha t^2 + c_1 t + c_2)\left[d\theta^2 + l(\theta)^2 d\phi^2\right]. \qquad (27)$$

Solving equations (5) to (14) using the space-time (27) one finds that $\psi = 0$ which means that no proper conformal vector fields exist. Conformal vector fields in this case are Killing vector fields which are given in equation (3).

## Case (vi):

Here, we have $A=(c_1t+c_2)^2$, $B=$constant, $R=2\alpha$ and $f(R)=\mathrm{e}^{\left[\frac{-\alpha}{c_1}\left(c_1\frac{t^2}{2}+c_2t\right)+c_3\right]}R+c_4$, where $c_1,c_2,c_3,c_4\in\Re\,(c_1\neq 0)$. The space-time (2) after suitable rescaling of $\theta$ and $\phi$ takes the form

$$ds^2=-dt^2+(c_1t+c_2)^2dr^2+\left[d\theta^2+l(\theta)^2d\phi^2\right]. \tag{28}$$

Again solving equations (5) to (14) with the help of the space-times (28) one finds that $\psi=0\Rightarrow$ no proper conformal vector fields exist in this case. Here conformal vector fields are Killing vector fields which are

$$\left.\begin{aligned}X^0&=c_5e^{c_1r}+c_6e^{-c_1r},\; X^1=\frac{-1}{(c_1t+c_2)}\left[c_5e^{c_1r}-c_6e^{-c_1r}\right]+c_7,\\ X^2&=c_8\cos\phi+c_9\sin\phi, X^3=\frac{l'}{l}(-c_8\sin\phi+c_9\cos\phi)+c_{10}\end{aligned}\right\}, \tag{29}$$

where $c_5,c_6,c_7,c_8,c_9,c_{10}\in\Re$. It is important to note that in this case the above space-time (28) admits six Killing vector fields. Four are given in equation (3) and other two Killing vector fields are $e^{c_1r}\frac{\partial}{\partial t}-\left(\frac{e^{c_1r}}{c_1t+c_2}\right)\frac{\partial}{\partial r}$, and $e^{-c_1r}\frac{\partial}{\partial t}+\left(\frac{e^{-c_1r}}{c_1t+c_2}\right)\frac{\partial}{\partial r}$.

## Case (vii):

Here, we have $A=t^4$, $B=t^{-1}$, $R=\frac{1}{2}\left[\frac{4\alpha t^3-7}{t^2}\right]$ and $f(R)=c_1\,\mathrm{e}^{\frac{-2\alpha t^3}{15}}R+c_2$, where $c_1,c_2\in\Re\,(c_1\neq 0)$. The space-time (2) takes the form

$$ds^2=-dt^2+t^4dr^2+t^{-1}\left[d\theta^2+l(\theta)^2d\phi^2\right]. \tag{30}$$

Solving equations (5) to (14) using the space-time (30) we find that $\psi=0$ which means that no proper conformal vector fields exist. Conformal vector fields in this case are also Killing vector fields and are given in equation (3).

## Case (viii):

In this case, we have $A=$constant, $B=t^2$, $R=2\left[\frac{\alpha-1}{t^2}\right]$ and $f(R)=c_1t^{\alpha-1}R+c_2$, where $c_1,c_2\in\Re\,(c_1\neq 0)$. The space-time (2) after an appropriate rescaling of $r$ takes the form:

$$ds^2 = -dt^2 + dr^2 + t^2\left[d\theta^2 + l(\theta)^2 d\phi^2\right]. \tag{31}$$

The Conformal vector fields in this case are:

$$\left.\begin{aligned} &X^0 = (c_3 r + c_4)t,\ X^1 = \left(\frac{t^2 + r^2}{2}\right)c_3 + c_4 r + c_5, \\ &X^2 = c_6 \cos\phi + c_7 \sin\phi,\ X^3 = \frac{l'}{l}(-c_6 \sin\phi + c_7 \cos\phi) + c_8 \end{aligned}\right\}, \tag{32}$$

where $c_3, c_4, c_5, c_6, c_7, c_8 \in \Re$. Conformal factor in this case is $\psi = (c_3 r + c_4)$. The above space-time (31) admits six linearly independent conformal vector fields in which four are Killing vector fields which are given in equation (3), one is proper homothetic vector field which is $t\frac{\partial}{\partial t} + r\frac{\partial}{\partial r}$ and one is proper conformal vector field. Proper conformal vector fields after subtracting the homothetic vector fields from equation (32) are:

$$X^0 = c_3 rt, X^1 = \left(\frac{t^2 + r^2}{2}\right)c_3, X^2 = 0,\ X^3 = 0. \tag{33}$$

## 3 Summary

In this paper, we studied conformal vector fields of some perfect fluid Kantowski-Sachs and Bianchi type III space-times in the f(R) theory of gravity using direct integration technique. Our purpose was twofold: firstly, we have found some perfect fluid solutions of Einstein field equations in the f(R) theory of gravity. Secondly, we have found conformal vector fields of the obtained solutions. From this study, we obtained the following results:

(a) The cases (i), (iii), (iv), (v), (vi) and (vii) in which conformal vector fields became the Killing vector fields. These are the space-times which are given in equations (21), (25), (26), (27). (28) and (30).

(b) In the cases (ii) and (viii), the above space-time (2) admits proper conformal vector fields which are given in equations (24) and (33). These space-times are given in equations (22) and (31).